\documentclass[
    reprint,
    aps,
    prl,
    amsmath,
    amssymb,
    superscriptaddress
]{revtex4-2}

\usepackage{graphicx}
\usepackage{dcolumn}
\usepackage{bm}
\usepackage{hyperref}
\usepackage[normalem]{ulem}
\usepackage{float}

\usepackage{xcolor}

\begin{document}

\title{Rapid Energy Dissipation by Colliding Waves in Strongly Magnetized Plasmas}

\author{Tianshu Wu}
\email{wuts25@mails.tsinghua.edu.cn}
\affiliation{
Department of Astronomy, Tsinghua University,
Beijing 100084, People's Republic of China
}

\author{Xinyu Li}
\email{xinyuli@mail.tsinghua.edu.cn}
\affiliation{
Department of Astronomy, Tsinghua University,
Beijing 100084, People's Republic of China
}

\author{Yangyang Cai}
\email{yangyang.cai@sjtu.edu.cn}
\affiliation{
Department of Astronomy, Tsinghua University,
Beijing 100084, People's Republic of China
}
\affiliation{
Tsung-Dao Lee Institute, Shanghai Jiao Tong University,
Shanghai 201210, China
}

\date{\today}

\begin{abstract}
Rapid dissipation of magnetic energy in highly magnetized environments around neutron stars and black holes is a key open question in high-energy astrophysics.
We develop a general kinetic picture of counter-propagating wave collisions in magnetized pair plasmas for arbitrary polarizations and find that magnetic energy can be dissipated on the wave-crossing timescale.
The two magnetohydrodynamical conditions on the field invariants, $I_1\equiv B^2-E^2>0$ and $I_2\equiv \mathbf{E}\cdot\mathbf{B}=0$, can be spontaneously violated during the collision.
Parallel electric fields develop to screen nonzero $I_2$ with
little energy loss, consistent with the evolution described by Force-Free Electrodynamics.
When magnetic dominance is lost, strong particle energization is triggered,  dissipating magnetic energy on the wave-crossing timescale.
This dynamical process yields a rapid dissipation channel of magnetic energy and
provides a kinetic pathway to high-energy emission.
\end{abstract}

\maketitle



\par
\indent\textit{Introduction.—}%
Rapid conversion of electromagnetic energy into energetic particles is
a central ingredient of high-energy activity in strongly magnetized
astrophysical systems, including magnetars and black-hole
magnetospheres \citep{ThompsonDuncan1995,BeloborodovThompson2007,
KaspiBeloborodov2017,LevinsonRieger2011,HirotaniOkamoto1998,Chen2018}.
It is required to explain emissions from black-hole coronae, pulsar-wind nebulae, solar flares, and other magnetically dominated environments.
Magnetic energy released by magnetospheric perturbations or instabilities is carried by Alfv\'en or magnetosonic waves which are subsequently dissipated to accelerate particles and power radiation.
Magnetic reconnection 
\citep{ThompsonDuncan1996,Lyutikov2003,Parfrey2013,Sironi2025,ZenitaniHoshino2001,Lyubarsky2005,SironiSpitkovsky2014,Guo2014} and
turbulent cascades
\citep{ThompsonBlaes1998,Li2019AlfvenFFE,Ripperda2021,GoldreichSridhar1995,Cho2005,Zhdankin2017,ComissoSironi2018} are two main mechanisms for magnetic energy dissipation.
Both mechanisms require waves to cross repeatedly before significant energy is dissipated \cite{Li2019AlfvenFFE}.
In magnetar magnetospheres, most wave energy is absorbed by the neutron star and dissipated internally to power a thermal afterglow \cite{Li2015} rather than being dissipated by magnetic reconnection or turbulence in the magnetosphere.
Moreover, fast transients in strongly magnetized systems occur on remarkably short timescales. Fast radio bursts exhibit millisecond durations and variability
\citep{lorimer2007,Thornton2013,Bailes2022}, while magnetar giant flares can rise on millisecond or submillisecond timescales
\citep{Mazets1979,Hurley1999,Hurley2005,Palmer2005}.
This timescale corresponds to a single wave crossing of the neutron star magnetosphere ($\sim 10R_{NS}/c$).
Such a short timescale requires a more efficient magnetic energy dissipation mechanism than magnetic reconnection and turbulence.

Wave dynamics and interactions in strongly magnetized environments are often described by Magnetohydrodynamics (MHD) and its high-magnetization approximation, Force-Free Electrodynamics (FFE).
Two conditions on the electromagnetic field invariants must be satisfied in the MHD/FFE description: magnetic dominance ($I_1\equiv B^2-E^2>0$) and the vanishing of the parallel electric field ($I_2\equiv\mathbf{E}\cdot \mathbf{B}=0$).

Recently, a new efficient dissipation mechanism has been proposed in which two waves with exactly anti-aligned magnetic fields collide and violate magnetic dominance \cite{Li2021AlfvenDissipation}.
The wave evolution deviates from the MHD/FFE description.
Plasma particles are accelerated to form a current sheet at the collision center that screens the excess electric field.
Magnetic energy from the waves is converted into plasma kinetic energy at the current sheet.
This mechanism can dissipate all the wave energy during a single collision when the wave magnetic fields are comparable to the background magnetic field.
Nevertheless, previous work \cite{Li2021AlfvenDissipation} considered only exactly anti-aligned wave collisions, a configuration that may be difficult to realize in astrophysical systems.
Realistic wave collisions can be more complicated, as parallel electric fields can also develop during the collision and accelerate particles.
Whether collisions between waves with arbitrary polarizations in magnetized plasmas can lead to efficient energy dissipation remains to be explored.

In this letter, we develop a general kinetic picture of wave collisions for various wave amplitudes and polarizations.
Using high-resolution 1D particle-in-cell (PIC) simulations, we perform first-principles calculations of nonlinear plasma responses.
We compare our results with FFE simulations and assess under what conditions FFE provides a good effective description of wave collisions.

\par
\indent\textit{Collision setup and MHD conditions.—}%
We consider two counter-propagating plane-wave packets along a uniform background magnetic field $\mathbf{B}_0=B_0\hat{\mathbf{x}}$ and colliding at $x=0$.
All quantities are independent of $y$ and $z$, so the electromagnetic
fields depend only on $x$ and time $t$. The background medium is a cold, charge-neutral electron--positron plasma with magnetization
$\sigma_0=B_0^2/(4\pi n_0mc^2)\gg1$, where $n_0=n_++n_-$ is the total particle number density and $m$ is the particle mass. 
Our wave mode has wavevector $\mathbf{k}\parallel\mathbf{B}_0$, carries no net current and propagates at nearly the speed of light like vacuum electromagnetic waves. 
It corresponds to the limit in which the Alfv\'en and fast modes become degenerate.

The field evolution follows Maxwell's equations
\begin{equation}
	\partial_t\mathbf{B}
	=
	-c\nabla\times\mathbf{E},
	\qquad
	\partial_t\mathbf{E}
	=
	c\nabla\times\mathbf{B}-4\pi\mathbf{J}.
\end{equation}
The collision becomes nontrivial when the current is determined by the nonlinear plasma response.
We denote the magnetic perturbations of the packets propagating along
$\pm\hat{\mathbf{x}}$ by $\delta\mathbf{B}_{\pm}$. They have equal
amplitude and form an angle $\theta$ in the transverse $y$--$z$ plane \footnote{For any two nonzero counter-propagating waves, a Lorentz boost along
$\mathbf{B}_0$ can transform their amplitudes to equal values.}.
Defining
$A=|\delta\mathbf{B}_{\pm}|/B_0
=|\delta\mathbf{E}_{\pm}|/B_0$, we adopt a convention symmetric about
the $z$ axis,
\begin{equation}
	\begin{pmatrix}
		\delta\mathbf{B}_{+}\\
		\delta\mathbf{B}_{-}
	\end{pmatrix}
	=
	AB_0
	\begin{pmatrix}
		\sin(\theta/2) & \cos(\theta/2)\\
		-\sin(\theta/2) & \cos(\theta/2)
	\end{pmatrix}
	\begin{pmatrix}
		\hat{\mathbf{y}}\\
		\hat{\mathbf{z}}
	\end{pmatrix}.
\end{equation}
The corresponding electric perturbations satisfy
$\delta\mathbf{E}_{\pm}
=\mp\hat{\mathbf{x}}\times\delta\mathbf{B}_{\pm}$.
The collision is therefore characterized by the wave amplitude $A$ and
the polarization angle $\theta$.

Before considering the plasma response, we first linearly superpose the two incoming packets.
At full overlap, the field invariants of this hypothetical current-free state are
\begin{eqnarray}
	I_1 &=& B_{\rm lin}^2 - E_{\rm lin}^2
	=B_0^2\left(1+4A^2\cos\theta\right),
	\\
	I_2 &=&\mathbf{E}_{\rm lin}\cdot\mathbf{B}_{\rm lin}= -2A^2B_0^2\sin\theta.
\end{eqnarray}
The subscript ``lin'' here denotes the linearly superposed state of the two waves. 
A parallel electric field appears unless the magnetic fields are aligned ($\theta=0^\circ$) or anti-aligned ($\theta=180^\circ$).
For $\cos\theta<0$, magnetic dominance is broken ($I_1<0$) when
\begin{equation}
    A>A_{\rm crit}(\theta)\equiv\frac{1}{2\sqrt{-\cos\theta}}.
\end{equation}
The simple linear estimate shows that wave collisions may violate either or both MHD conditions and therefore depart from the MHD description.

FFE is the strong-magnetization limit of MHD, in which plasma inertia is neglected and the Lorentz force density $\rho_e\mathbf{E}+\mathbf{J}\times\mathbf{B}/c$ vanishes.
The standard force-free current prescription preserves $I_2=0$ with $\mathbf{J}_{\rm FFE}\cdot\mathbf{E}=0$
\citep{Komissarov2002}, allowing the current to reorganize the fields
without transferring electromagnetic energy to the plasma. 
However, $I_1>0$ is not guaranteed by the FFE evolution and may be dynamically violated.
Kinetic plasma responses must therefore be considered to accurately capture the wave-collision dynamics.

\indent\textit{Numerical Method.—}We perform one-dimensional special-relativistic particle-in-cell
simulations using \textsc{Entity} \citep{Hakobyan2026Entity}. The
simulations follow a cold, charge-neutral electron--positron plasma with magnetization $\sigma_0=25$, initialized with two counter-propagating square wave packets specified by the amplitude $A$ and polarization angle $\theta$ defined above. We set $c/\omega_p=1$, where  
$\omega_p=(4\pi e^2n_0/m)^{1/2}$ is the plasma frequency, and resolve the skin depth with 30 grid cells. The simulations use $6\times10^4$ grid cells and 512
particles per cell, with $\lambda\omega_p/c=820$.
For comparison, we evolve the same initial electromagnetic
configuration using the FFE code of
\citet{Li2019AlfvenFFE}, which employs a fifth-order WENO method and
restores the MHD conditions by damping the parallel electric
field when $I_2\neq0$ and rescaling $\mathbf{E}$
when $I_1<0$.
Full numerical details and convergence tests are provided in the Supplemental Material.

\begin{figure*}[t]
	\centering
	\includegraphics[width=0.98\textwidth]{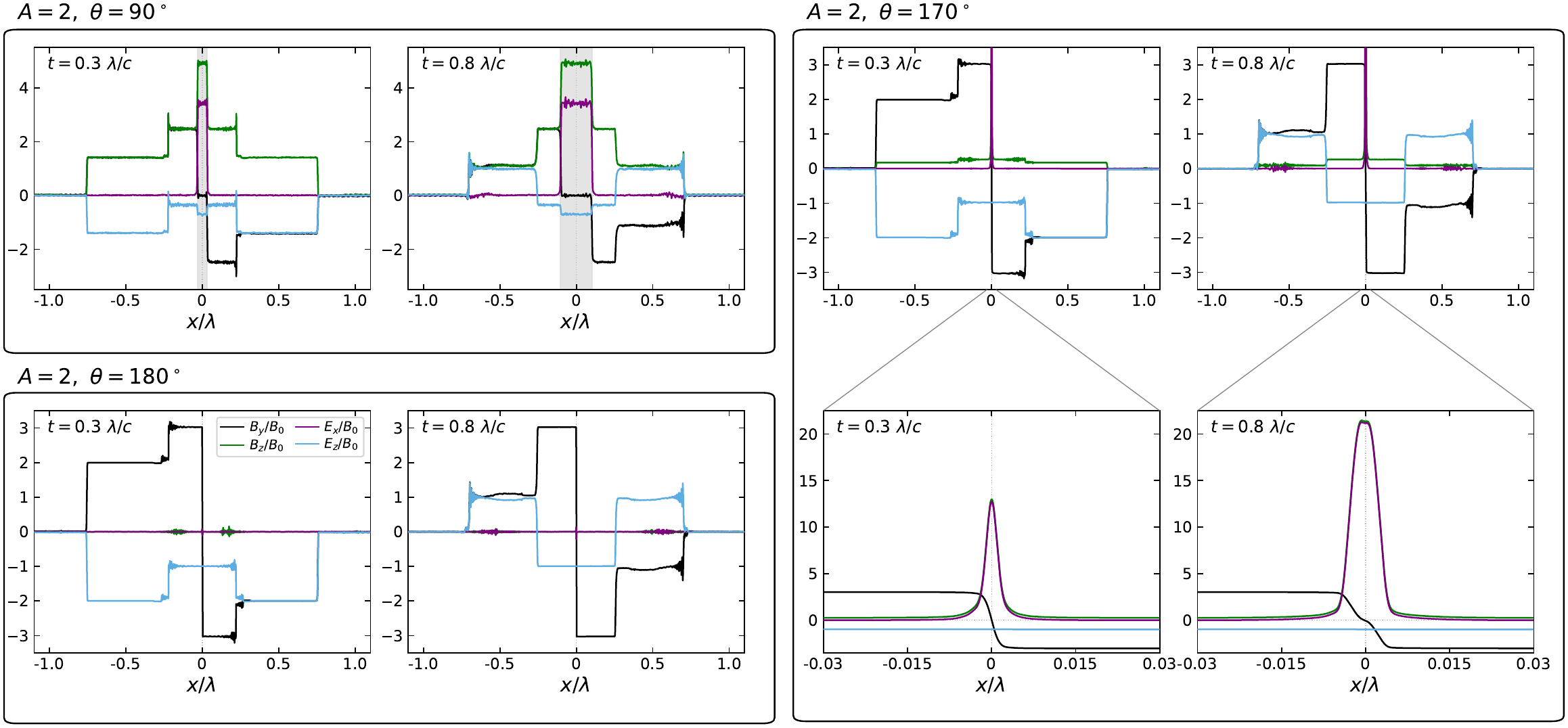}
	\caption{
	Evolution of the normalized electromagnetic fields for $A=2$ at three
	representative polarization angles. The left block shows
	$\theta=90^\circ$ (top) and exact anti-alignment,
	$\theta=180^\circ$ (bottom), at $ct/\lambda=0.3$ and $0.8$.
	The right block shows the oblique case $\theta=170^\circ$ at the same
	times, with the global profiles in the top row and a kinetic-scale zoom
	around the collision center in the bottom row. Black, green, magenta,
	and blue curves denote $B_y/B_0$, $B_z/B_0$, $E_x/B_0$, and $E_z/B_0$,
	respectively. 
	}
	\label{fig:field_evolution}
\end{figure*}

\par
\indent\textit{Representative field responses.—}%
Fig.~\ref{fig:field_evolution} shows the evolution of electromagnetic fields for $A=2$ and three different polarization angles $\theta=90^\circ,170^\circ$ and $180^\circ$.
In all three cases, waves are reflected from the collision center with the same polarization and form a magnetic jump. 

The upper left panels show the perpendicular case, $\theta=90^\circ$.
Magnetic dominance is preserved throughout the evolution, but the linear analysis predicts the development of a parallel electric field.
$E_x$ is generated to restore $I_2=0$ in the gray regions.
During the collision, the gray regions expand from the center at the characteristic speeds of the outgoing current-carrying Alfv\'en waves
\citep{Komissarov2002}, whose $x$-components are
\begin{equation}
	v_{A,x}^{\pm}
	=
	c\frac{
		(\mathbf{E}\times\mathbf{B})_x
		\pm B_x\sqrt{B^2-E^2}
	}{
		B^2
	}.
	\label{eq:ffe_alfven_characteristic}
\end{equation}

\begin{figure}[t]
	\centering
	\includegraphics[width=1.00\columnwidth]{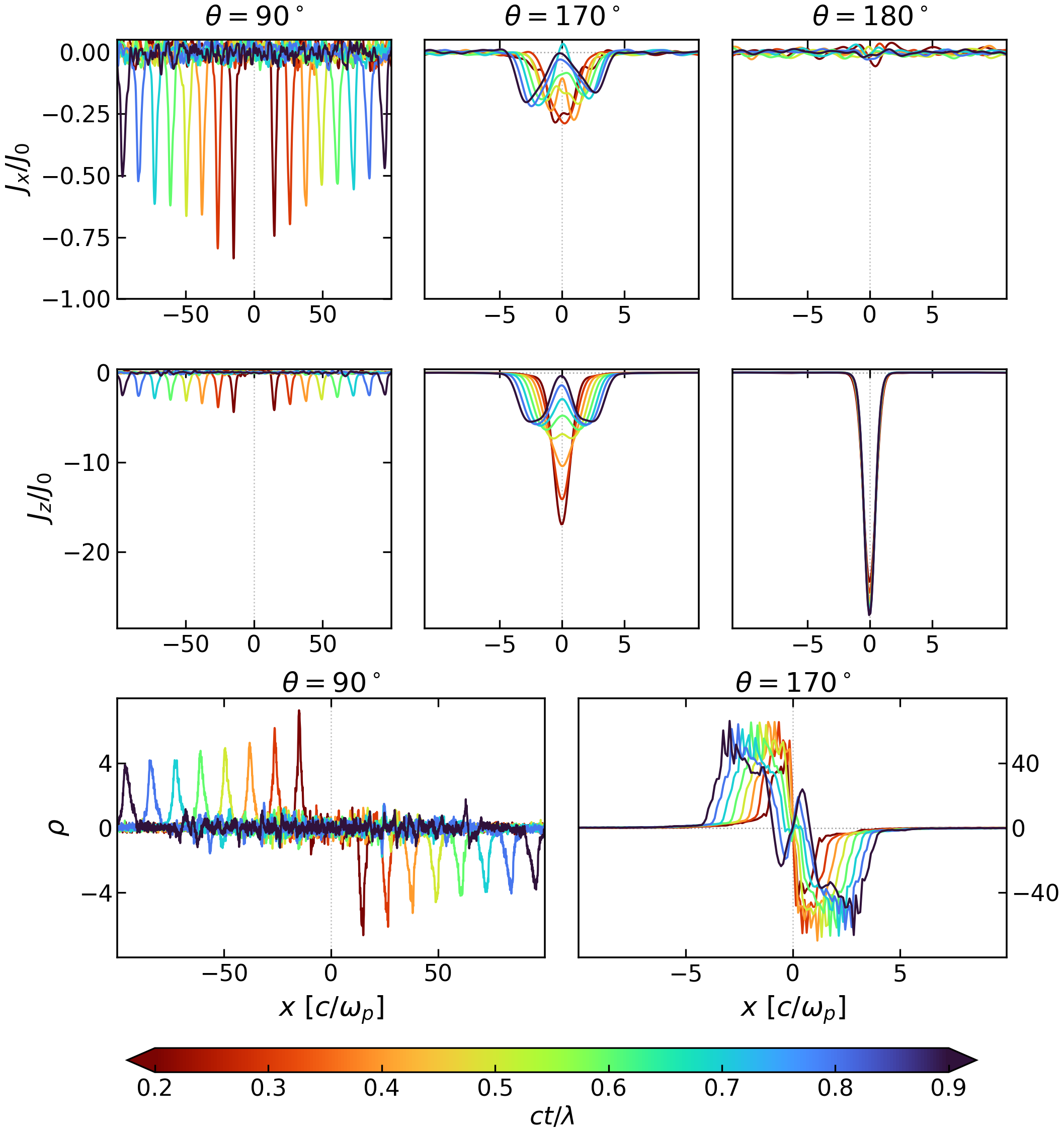}
	\caption
    {
	Evolution of the current and charge-density structures for $A=2$. The top and middle rows show $J_x/J_0$ and $J_z/J_0$ for $\theta=90^\circ$, $170^\circ$, and $180^\circ$, where $J_0=B_0\omega_p/(4\pi)$ is the current-density normalization. The bottom row shows the charge density $\rho$ for $\theta=90^\circ$ and $170^\circ$. Colors denote time from $ct/\lambda=0.2$ to $0.9$, and distance is measured in units of $c/\omega_p$.
	}
	\label{fig:currents_charge}
\end{figure}

The lower left panels in Fig.~\ref{fig:field_evolution} show the exact anti-aligned case ($\theta=180^\circ$) where $I_2=0$ during the collision, but magnetic dominance is violated.
The results are consistent with \citet{Li2021AlfvenDissipation}.
Waves are reflected with amplitude $A-1$ and form a current sheet at the collision center.
The right panels of Fig.~\ref{fig:field_evolution} show the oblique case
$\theta=170^\circ$, where both MHD conditions are violated.
The field evolution combines features of both previous cases.
There is an expanding region of $E_x$ together with $B_z$ similar to the perpendicular case, and $B_y$ exhibits a magnetic jump similar to the anti-aligned case.
The reflected waves have smaller amplitudes than the incoming waves, 
indicating wave dissipation associated with the simultaneous violation of the two MHD conditions.

Direct comparisons between PIC and FFE simulations are presented in the Supplemental Material.
The field evolutions are indistinguishable in all tested cases where magnetic dominance is not violated.
When only the parallel electric field needs to be screened, the kinetic response of the plasma develops an induced current that is well described by FFE, and the macroscopic evolution remains effectively force-free without dissipation.

\begin{figure}[t]
	\centering
	\includegraphics[width=0.98\columnwidth]{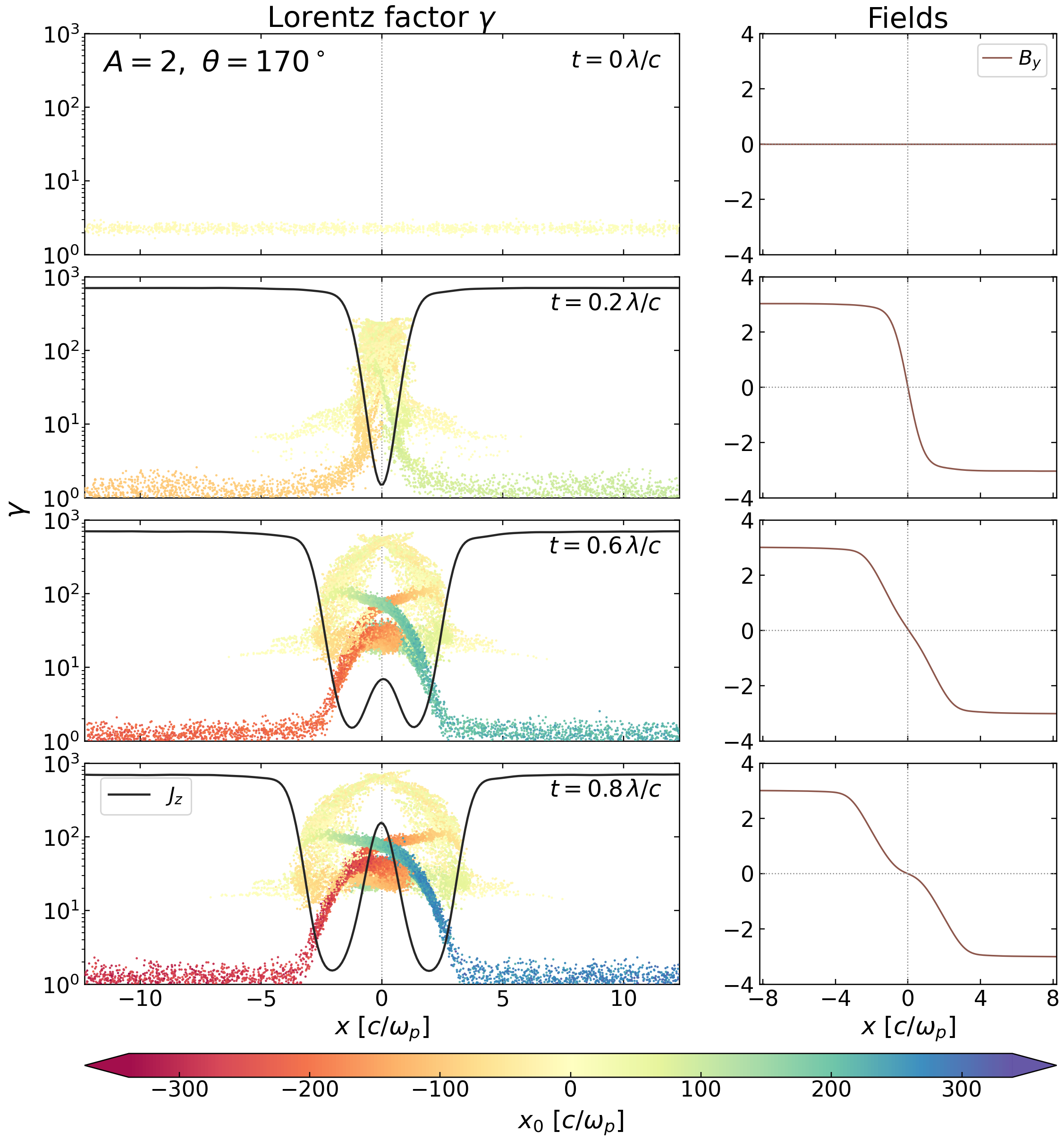}
	\caption{
	Particle energization and current support for $A=2$ and
	$\theta=170^\circ$. The left panels show the particle Lorentz factor
	$\gamma$ at $ct/\lambda=0$, $0.2$, $0.6$, and $0.8$, colored by the
	initial position $x_0$. The corresponding $B_y$ profiles are shown on
	the right. Black curves show $J_z$, rescaled for visualization. 
    Energetic particles progressively populate the outward current branches, while the opposite $v_z$ and opposite charges of the two species yield contributions of the same sign to $J_z$.
	}
	\label{fig:particle_evolution}
\end{figure}

\par
\indent\textit{Current dynamics and particle energization.—}%
Fig.~\ref{fig:currents_charge} shows the evolution of current and charge density during the wave collision.
Current sheets of skin-depth width are induced when the MHD conditions are violated, although there is no current in the incoming waves.
For the perpendicular case, the $x$-component of the current supports the induced $E_x$ to restore $I_2=0$, causing positive and negative charges to separate and move outward from the center with the Alfv\'en characteristic speed given by Eq.~\eqref{eq:ffe_alfven_characteristic}.
For the anti-aligned case, a strong $J_z$ is induced by the acceleration of plasma particles by the excess $E_z$, with no significant $J_x$ or charge separation observed.
$J_z$ remains stationary during the collision to keep $I_1$ close to zero.
For the oblique case ($\theta=170^\circ$), a strong $J_z$ initially develops at the center and later splits into two identical currents with separated charges propagating outward as Alfv\'en waves.
A finite polarization offset therefore converts the stationary central current layer into a dynamically evolving current system.
The split current sheets retain the same width as the original one. The initial magnetic jump in $B_y$ transforms into two steps, each with the same width but half the amplitude, linked by an expanding $B_y=0$ region at the center.
Accelerated particles can then escape the magnetic jumps either ahead of the Alfv\'en waves or into the central $B_y=0$ region.

The current sheet and magnetic jump produce sustained particle acceleration only when magnetic dominance is violated, while the interaction remains essentially elastic when only $I_2\neq0$. When $I_1>0$, the system remains hyperbolic, and energy can be carried away by outgoing characteristic modes without irreversible transfer to the plasma. For the $\theta=90^\circ$ case, although some initial particle acceleration accompanies the development of $E_x$, no sustained acceleration is observed. Particles encounter only a static, spatially rotating magnetic field and complete multiple gyro-orbits before being elastically reflected with no net energy gain.

When $I_1<0$, however, the system loses hyperbolicity. 
Particles traversing the current layer become unmagnetized near the
reversal of the transverse magnetic-field component,
and their trajectories follow Speiser orbits \citep{Speiser1965,Li2021AlfvenDissipation}. 
During this motion, they gain energy from the electric field and are
subsequently ejected from the layer by the magnetic pressure gradient near its edges. This process occurs in the comoving frame of a propagating magnetic jump
after it splits and moves outward, as in the $\theta=170^\circ$ case,
just as it does in the stationary case at $\theta=180^\circ$.
Fig.~\ref{fig:particle_evolution} shows the particle energy--position
phase space at four different times, colored by their initial positions $x_0$ and overlaid with the $J_z$ profiles in black. The corresponding $B_y$ profiles of the magnetic jumps are shown on the right.
Particles starting near the collision center acquire high energies
during the initial relaxation that screens the excess electric field.
As the current branches propagate outward as Alfv\'en modes, plasma
particles are successively accelerated to $\gamma\sim\sigma_0$.

The rate of particle energization is controlled by two characteristic timescales:
the transit time across the current sheet, $\tau_t\sim\Delta/c\sim1/\omega_p$, where $\Delta$ is the sheet thickness,
and the gyro-time in the perpendicular magnetic field,
$\tau_g\sim\gamma mc/(eB_0)$, where $\gamma$ is the particle Lorentz factor. For low-energy particles entering the layer, $\tau_t/\tau_g\sim\sqrt{\sigma_0}$.
In a strongly magnetized system, $\tau_t\gg\tau_g$, so particles complete
multiple gyro-orbits before crossing the layer, and the net work done by
the electric field averages to nearly zero.
In the presence of a perpendicular accelerating electric field, particles
entering the current sheet are first accelerated to high $\gamma$.
As $\gamma$ increases, the gyro-period lengthens, allowing particles
to follow Speiser orbits while remaining confined within the magnetic jump.
They are ejected when $v_z$ changes sign, with $E_z$ doing net work
during the motion. This energization drives the system toward
$\tau_t\sim\tau_g$, which self-consistently sets the current-sheet width
$\Delta$.
Since the accelerating field scales as
$E_z\propto A\sin(\theta/2)$, the characteristic particle Lorentz factor
increases with wave amplitude and polarization angle and is largest at
exact anti-alignment ($\theta=180^\circ$).

\par
\begin{figure}[t]
	\centering
	\includegraphics[width=\columnwidth]{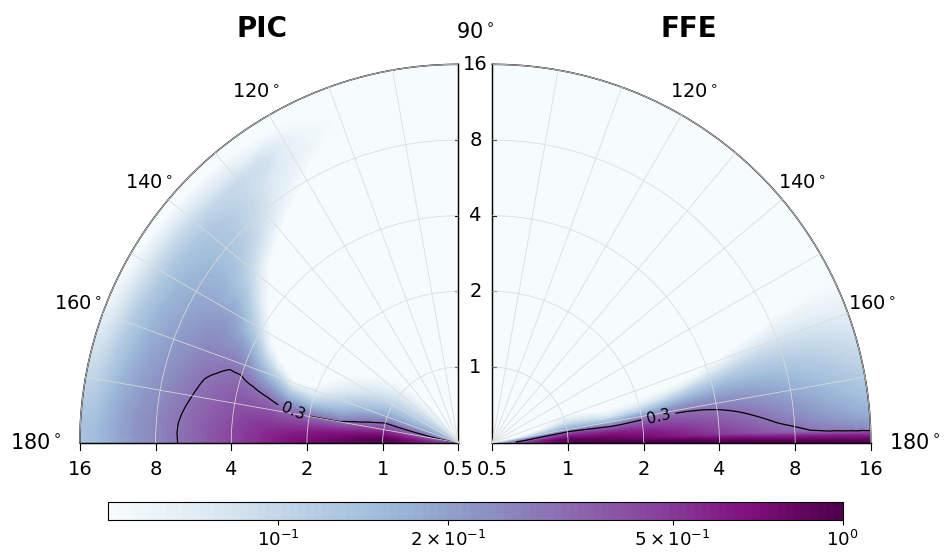}
	\caption{
		Dissipation rate $f$ across wave amplitude $A$ and polarization angle
		$\theta$ in PIC (left) and FFE (right), measured from the decline of
		the normalized free-wave energy $U_{\rm wave}(t)/U_{\rm wave}(0)$.
		The radial coordinate $A$ and the color scale are logarithmic.
		The black contour marks $f=0.3$ in the PIC results.
	}
	\label{fig:dissipation_map}
\end{figure}

\indent\textit{Global energetic response.—}%
We define the wave free energy as
\begin{equation}
	U_{\rm wave}(t)
	=
	\int \frac{E^2+B^2-B_0^2}{8\pi}\,dx,
	\label{eq:wave_energy}
\end{equation}
where the energy of the uniform background magnetic field is subtracted.
The dissipation rate $f$ is obtained from a linear fit to the nearly steady decline of $U_{\rm wave}(t)/U_{\rm wave}(0)$. 
Fig.~\ref{fig:dissipation_map} maps this dissipation rate across the amplitude--polarization parameter space for both PIC and FFE simulations.
No measurable dissipation is found for $\theta<90^\circ$, consistent with magnetic dominance being preserved, and this range is therefore not shown.
In the PIC simulations, dissipated field energy powers particle acceleration, with total energy conserved to numerical accuracy.
Ideal FFE transfers no energy through $\mathbf{J}\cdot\mathbf{E}$.
Its apparent energy removal instead arises from rescaling the electric field when $E>B$ to restore magnetic dominance. 

The PIC results in Fig.~\ref{fig:dissipation_map} reveal a broad region of significant energy dissipation, showing the parameter space in which particle acceleration and potentially associated radiation can occur. 
For fixed amplitude, the dissipation is strongest near exact anti-alignment \footnote{Here we plot the dissipation rate $f$. The total dissipated energy increases monotonically with $A$. However, $f$ is normalized by the initial wave energy, which scales as $A^2$, so its dependence on $A$ becomes nonmonotonic and develops peaks and valleys. The same behavior is present in the anti-aligned case, for which an analytical result is available.}.
With increasing amplitude, significant dissipation extends to smaller
polarization angles. A fitting formula for the dependence of $f$
on $A$ and $\theta$ is provided in the Supplemental Material.
In comparison, the FFE results deviate significantly from the PIC results at large amplitude. The dissipation rate is overestimated near anti-alignment and underestimated at small polarization angles.

\par
\indent\textit{Discussion and Conclusions.—}%
We investigate the general collision of counter-propagating non-current-carrying waves with relative polarization offsets in a strongly magnetized pair plasma. We perform systematic one-dimensional PIC simulations across the amplitude--polarization parameter space and compare with FFE simulations. 
Our simulations demonstrate that polarization geometry governs the kinetic response and energy transfer. Our main conclusions are as follows:
\begin{enumerate}
\item Unlike linear propagation in vacuum, where waves cross each other, the waves are reflected with the same polarization, forming a magnetic jump at the center.

\item Violations of the two MHD conditions are common and have different consequences. 
Nonzero $\mathbf{E}\cdot\mathbf{B}$ is screened by an induced electric field parallel to the background magnetic field, leading to charge separation and launching outgoing current-carrying Alfv\'en waves without significant wave dissipation. Fast dissipation occurs only when magnetic dominance is lost and is most efficient when the two waves are anti-aligned.

\item The dissipated wave energy is converted into the kinetic energy of plasma particles when they stream into the magnetic jump.

\item FFE provides an accurate description only when magnetic dominance is not violated.

\end{enumerate}

Collisions of counter-propagating waves are expected to occur
widely in strongly magnetized astrophysical environments including magnetar magnetospheres and black-hole coronae. 
Our results show that strong dissipation of wave energy within a single wave-crossing timescale can happen during the collision.
Particles are accelerated to energies of the order of the magnetization
and may power high-energy radiation.
In the neutron star magnetosphere, the characteristic wave-crossing timescale is $\sim 10R_{NS}/c=1$~ms.
Collisional dissipation of waves provides a possible mechanism for fast transient high-energy phenomena like magnetar bursts and fast radio bursts which occur on millisecond timescales.

The presented simulations are limited to 1D non-current-carrying waves in a uniform background. In two dimensions, the tearing instability may trigger magnetic
reconnection and provide an additional dissipation channel.
Our kinetic simulations also neglect radiative losses of the accelerated particles.  
Synchrotron losses may limit the energy of ultra-relativistic particles in the current sheet, and Compton scattering can produce energetic photons that subsequently convert to pairs.
These radiative processes may power X-ray bursts similar to those produced by radiative magnetic reconnection \citep{beloborodov2021} or gamma-ray flares \citep{Giannios2009}.
Future work should therefore extend our calculations to multidimensional PIC simulations with more realistic settings in magnetar or black-hole magnetospheres.




\bibliographystyle{apsrev4-2}
\bibliography{AFD-revtex}

\clearpage

\begin{center}
{\large\bf Supplemental Material}\\[0.5em]
\end{center}

\section{\label{sec:supp-methods}Numerical Methods and Convergence Tests}
\label{app:convergence_test}

\subsection{Particle-in-cell simulations}
We perform one-dimensional special-relativistic particle-in-cell (PIC)
simulations using \textsc{Entity}, a hardware-agnostic PIC code for
astrophysical plasmas that evolves the Vlasov--Maxwell system
\citep{Hakobyan2026Entity}. The simulations follow a cold,
charge-neutral electron--positron plasma with magnetization
$\sigma_0=25$ in a uniform background magnetic field
$\mathbf{B}_0=B_0\hat{\mathbf{x}}$.

Two square wave packets of equal amplitude are initialized to
counter-propagate along the background magnetic field. Their magnetic and electric perturbations follow the
amplitude--polarization convention defined in the main text. The collision is
characterized by the wave amplitude $A$ and the polarization angle
$\theta$. Electrons and positrons are initialized with the same local
drift velocity
$
	\mathbf{v}_{\rm d}
	=
	c\frac{\mathbf{E}\times\mathbf{B}}{B^2},
$
so that each isolated packet initially carries no net plasma current.

The computational domain extends over
$-1000\leq x\leq1000$ and is resolved by $6\times10^4$ grid cells,
corresponding to a grid spacing $\Delta x=1/30$. Periodic boundary
conditions are imposed on both particles and electromagnetic fields, and
we analyze only the first collision of the two wave packets. The plasma has a uniform total number density $n_0=n_++n_-$, which
defines the reference plasma frequency
$\omega_p=(4\pi e^2n_0/m)^{1/2}$.
We set the plasma skin depth to $c/\omega_p=1$, which is resolved by
30 grid cells. For $\sigma_0=25$, the cyclotron frequency satisfies
$\omega_B/\omega_p=\sqrt{\sigma_0}=5$, giving the characteristic
cyclotron length $c/\omega_B=0.2$. The particle distribution is sampled
with 512 computational particles per cell in total, equally divided
between electrons and positrons.

\begin{figure}
	\centering
	\includegraphics[width=1.0\columnwidth]{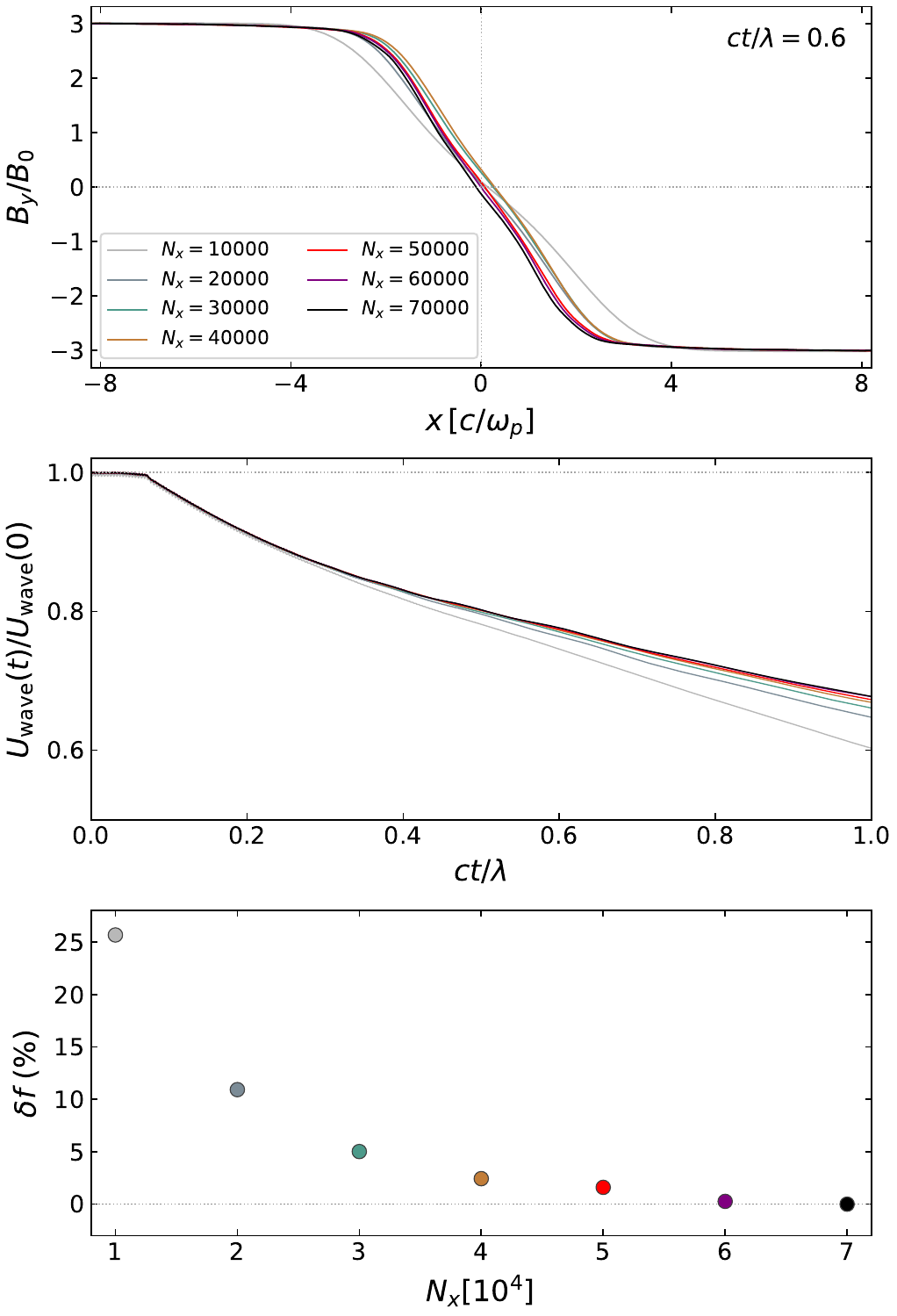}
	\caption{
	Spatial resolution convergence test for the case $A=2$ and
	$\theta=170^\circ$. The top panel shows $B_y/B_0$ at
	$ct/\lambda=0.6$ for different numbers of grid cells $N_x$.
	The middle panel shows the corresponding evolution of the normalized
	free-wave energy $U_{\rm wave}(t)/U_{\rm wave}(0)$.
	The bottom panel shows the dissipation rate error $\delta f$ as a
	function of $N_x$.
	}
	\label{fig:resolution_convergence}
\end{figure}

\begin{figure}
	\centering
	\includegraphics[width=1.0\columnwidth]{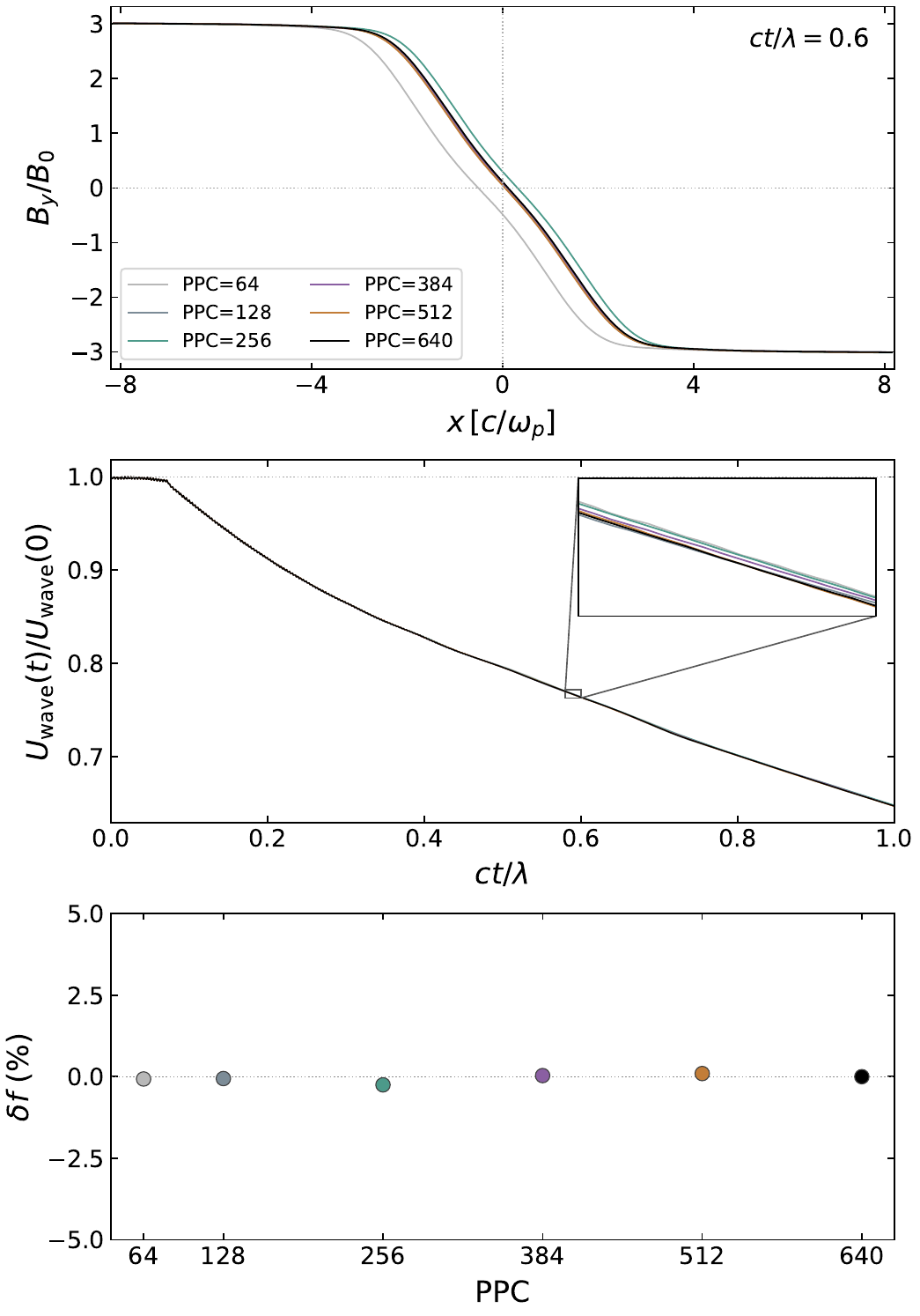}
	\caption{
	Particle number convergence test for the case $A=2$ and
	$\theta=170^\circ$. The top panel shows $B_y/B_0$ at
	$ct/\lambda=0.6$ for different numbers of particles per cell (PPC).
	The middle panel shows the corresponding evolution of the normalized
	free-wave energy $U_{\rm wave}(t)/U_{\rm wave}(0)$, with the inset
	magnifying the boxed region. The bottom panel shows the
	dissipation rate error $\delta f$ as a function of PPC.
	}
	\label{fig:ppc_convergence}
\end{figure}

\subsection{Force-free simulations}

For comparison, we evolve the same initial electromagnetic configuration using FFE. The numerical method follows
\citet{Li2019AlfvenFFE} and employs a flux-conservative scheme that is
third order in time and fifth order in space, based on the weighted
essentially non-oscillatory (WENO) method.

The numerical evolution includes explicit treatments to preserve the
MHD conditions. Hyperbolic divergence cleaning is used to control
violations of $\nabla\cdot\mathbf{B}=0$. Small deviations from
$\mathbf{E}\cdot\mathbf{B}=0$ are suppressed by a modified force-free
current that damps the parallel electric field over a finite timescale.
When magnetic dominance is violated ($E>B$), the electric field is rescaled
according to
$
	\mathbf{E}
	\rightarrow
	\sqrt{\frac{B^2}{E^2}}\,
	\mathbf{E},
$
thereby restoring $E\leq B$. Further details of the numerical
implementation are given in \citet{Li2019AlfvenFFE}.

\subsection{Convergence Tests}

We examine numerical convergence using the representative case
$A=2$ and $\theta=170^\circ$. We use three diagnostics: the profile of
the reversing magnetic field $B_y$ at $t=0.6\lambda/c$, the evolution
of the free-wave energy, and the dissipation-rate error $\delta f$. 
We define $\delta f=(f-f_{\rm ref})/f_{\rm ref}\times100\%$, where $f_{\rm ref}$ is obtained from the highest-resolution run in each test.

Fig.~\ref{fig:resolution_convergence} shows the spatial resolution
test with $N_x=\{1,2,3,4,5,6,7\}\times10^4$, while all other parameters are held fixed. 
Increasing $N_x$ sharpens the $B_y$ reversal and brings the wave free energy curves toward a common evolution. 
At low resolution, the kinetic layer is numerically broadened and the dissipation rate is overestimated. 
The field profiles and energy evolution converge systematically, while $\delta f$ decreases steadily with increasing $N_x$, demonstrating spatial convergence.

Fig.~\ref{fig:ppc_convergence} shows the particle number test with
${\rm PPC}=\{64,128,256,384,512,640\}$. 
Apart from the lowest PPC run,
the $B_y$ profiles are close and the normalized wave free energy curves nearly overlap. 
The dissipation-rate errors remain below approximately $1\%$, with small nonmonotonic variations arising from particle statistics. 
The dissipation rate is therefore considerably less sensitive to PPC than to spatial resolution and is well converged over the tested range.

\begin{figure*}
	\centering
	\includegraphics[width=0.95\textwidth]{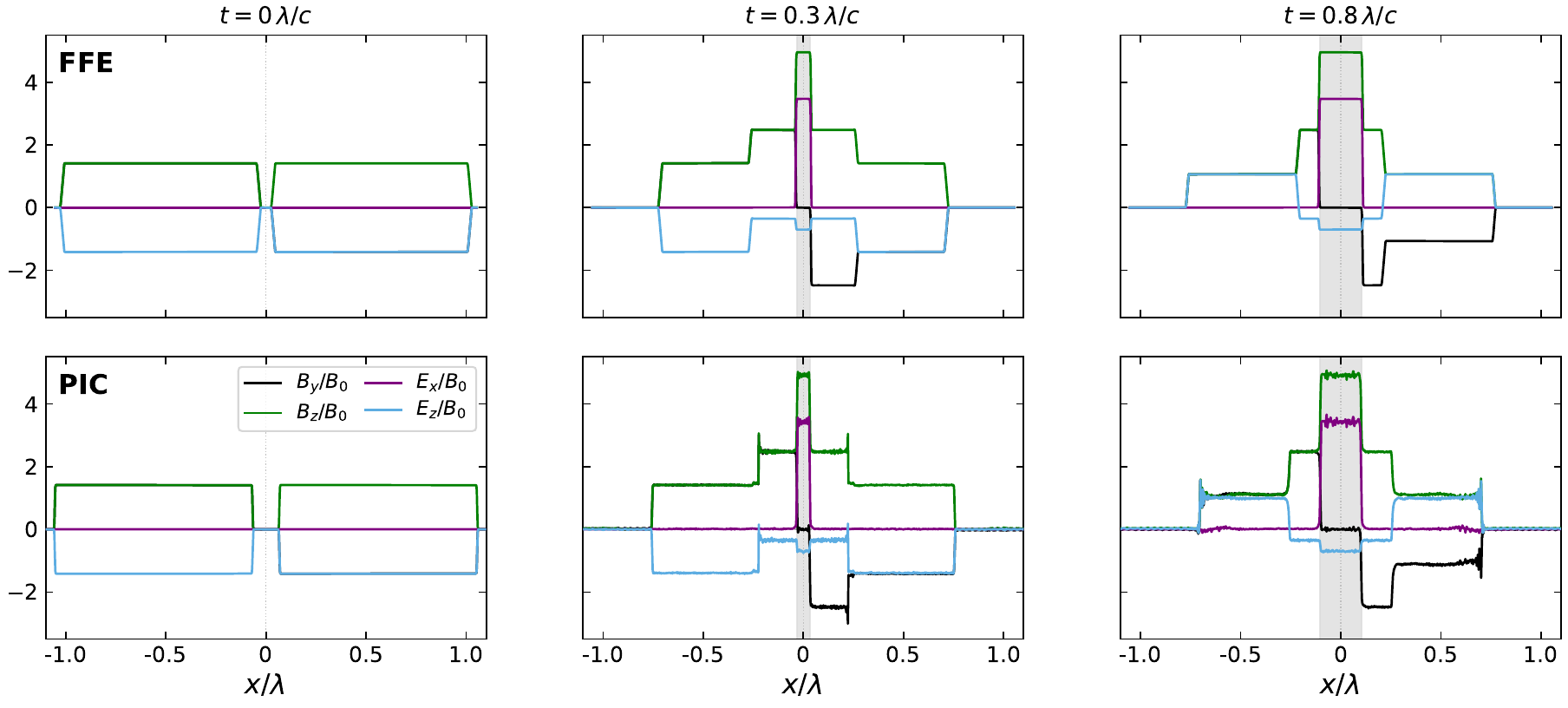}
	\caption{
	Comparison of the field evolution for $A=2$ and $\theta=90^\circ$ in
    FFE (top) and PIC (bottom) at $ct/\lambda=0$, $0.3$, and $0.8$.
    Black, green, magenta, and blue curves show $B_y/B_0$, $B_z/B_0$,
    $E_x/B_0$, and $E_z/B_0$, respectively. Gray shading marks the region between the outgoing field structures.
	}
	\label{fig:ffe_com_A2angle90}
\end{figure*}

\begin{figure*}
	\centering
	\includegraphics[width=0.95\textwidth]{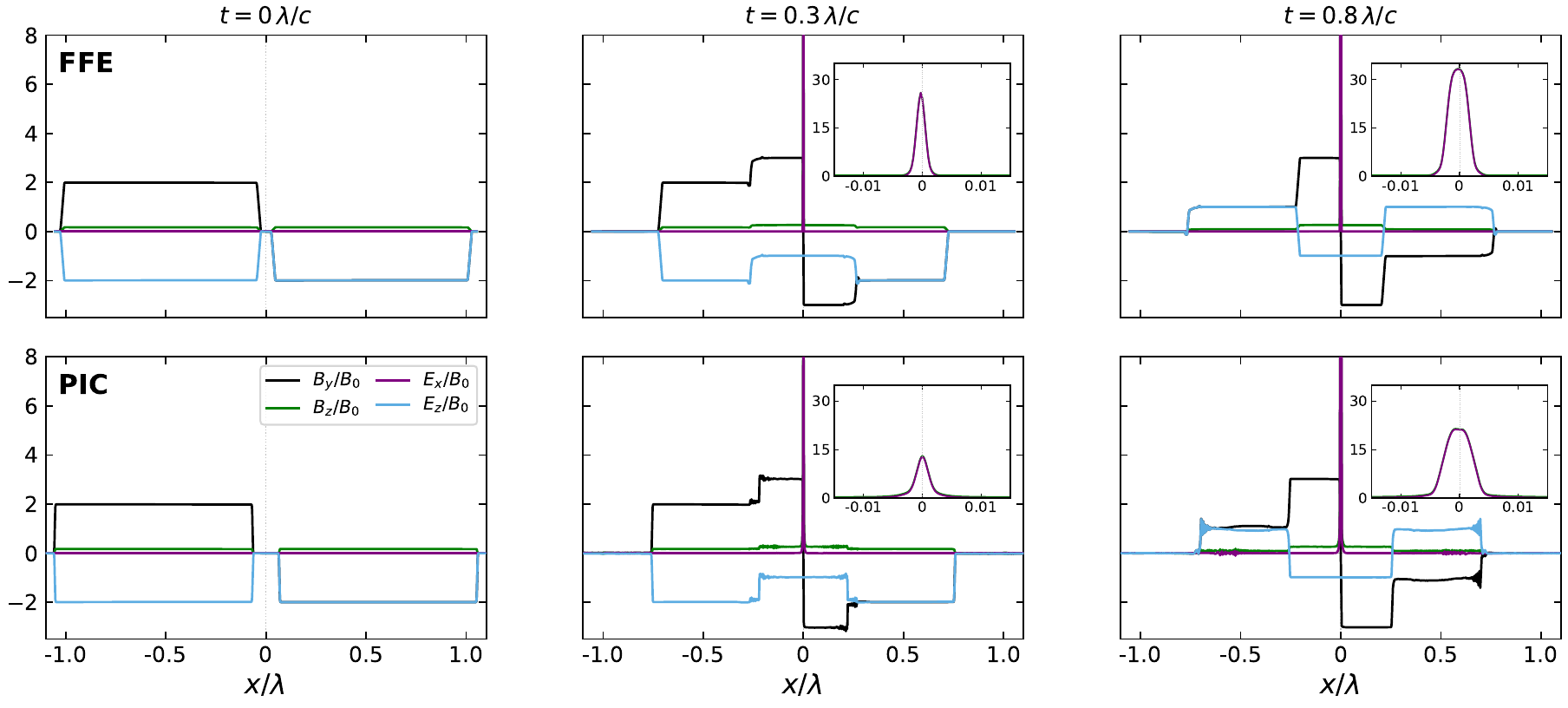}
	\caption{
	Comparison of the field evolution for $A=2$ and $\theta=170^\circ$ in
    FFE (top) and PIC (bottom) at $ct/\lambda=0$, $0.3$, and $0.8$.
    Black, green, magenta, and blue curves show $B_y/B_0$, $B_z/B_0$,
    $E_x/B_0$, and $E_z/B_0$, respectively. 
	}
	\label{fig:ffe_com_A2angle170}
\end{figure*}

\begin{figure*}
	\centering
	\includegraphics[width=0.95\textwidth]{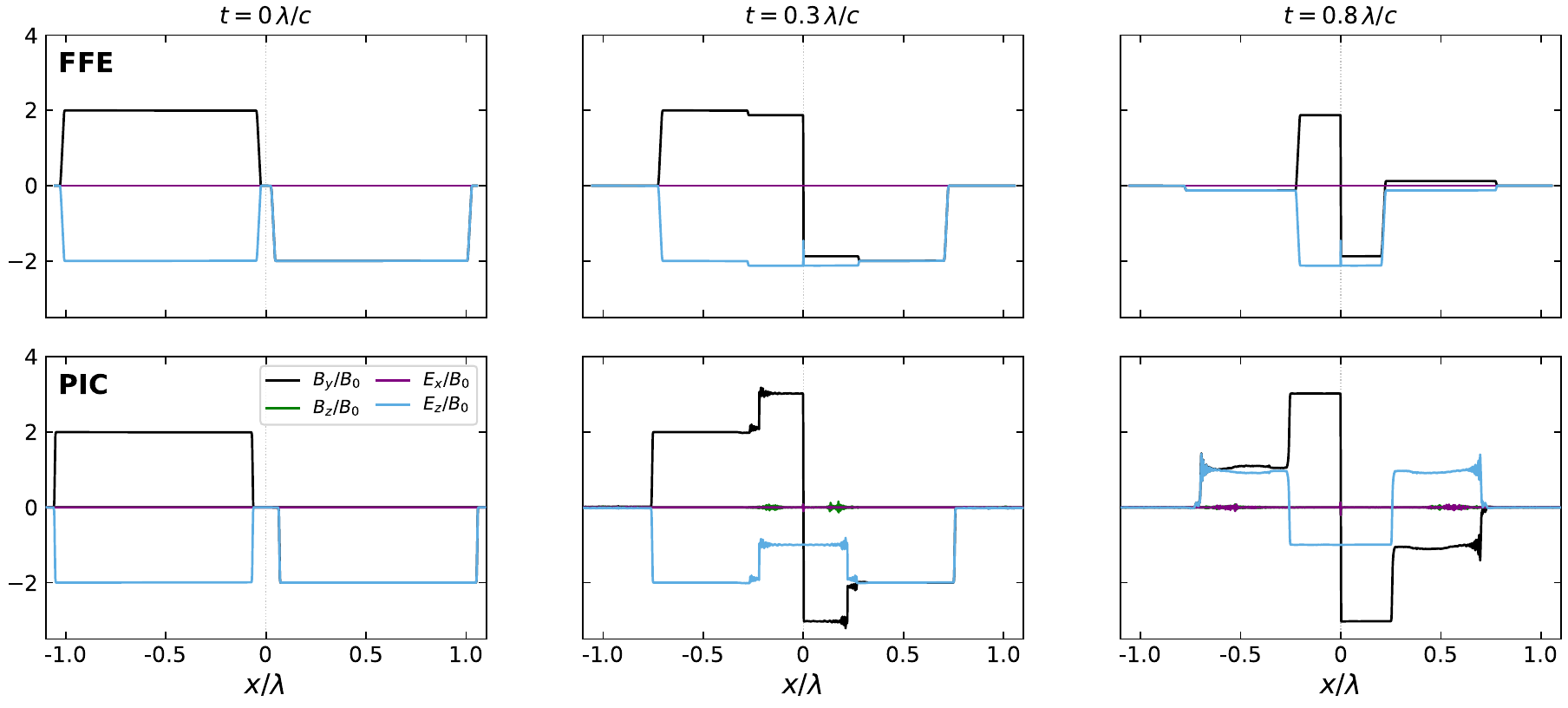}
	\caption{
	Comparison of the field evolution for $A=2$ and $\theta=180^\circ$ in
    FFE (top) and PIC (bottom) at $ct/\lambda=0$, $0.3$, and $0.8$.
    Black, green, magenta, and blue curves show $B_y/B_0$, $B_z/B_0$,
    $E_x/B_0$, and $E_z/B_0$, respectively. 
	}
	\label{fig:ffe_com_A2angle180}
\end{figure*}

\section{Comparison between PIC and FFE simulations}
\label{app:transition}

\begin{figure}
	\centering
	\includegraphics[width=0.95\columnwidth]{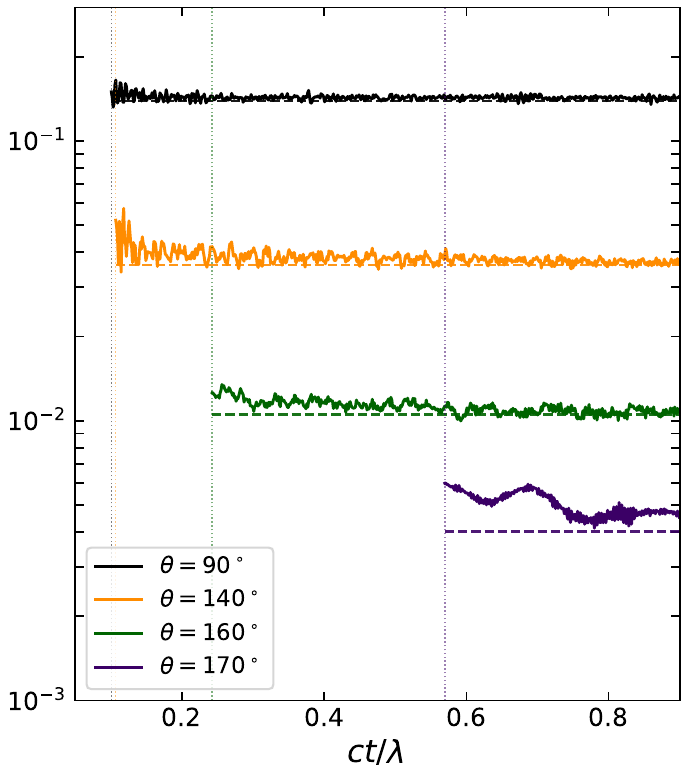}
	\caption{
        Comparison between the propagation speed of the outward-moving $J_z$ branch toward $+x$ and the local force-free Alfv\'en characteristic for $A=2$ and $\theta=90^\circ$, $140^\circ$, $160^\circ$, and $170^\circ$. Solid curves show the corresponding outward Alfv\'en characteristic speed, $v_{A,x}^{\rm out}/c$, evaluated from the local PIC fields near the current branch. Horizontal dashed lines show the measured effective branch speeds $v_{\rm br}/c$.
	}
	\label{fig:alfven_branch_speed}
\end{figure}

Figs.~\ref{fig:ffe_com_A2angle90} to \ref{fig:ffe_com_A2angle180} compare the electromagnetic fields obtained from PIC and FFE simulations for $\theta=90^\circ$, $170^\circ$, and $180^\circ$.
The fields obtained from the PIC and FFE simulations are identical for the perpendicular case, but start to deviate at $\theta=170^\circ$, showing different $B_z$ at the collision center. 
For $\theta=180^\circ$, FFE fails to reproduce the PIC results, as the
rescaling of the electric field removes too much wave energy from the
system, and little wave energy is reflected from the collision.

In general, we find that FFE can reproduce the PIC results when magnetic
dominance is not violated. The self-consistent FFE current provides an accurate description of the
plasma response when parallel electric fields tend to develop but
magnetic dominance is preserved. The usual rescaling of the excess electric field is not physical.

\section{Speed of the current branches}
\label{app:alfven_vel}
Fig.~\ref{fig:alfven_branch_speed} compares the current-branch speeds
measured from PIC simulations (dashed lines) with the local force-free
Alfv\'en characteristic speeds (solid lines) for different polarization
angles. We measure the branch speeds from the positions of the local
$J_z$ extrema after the two current branches separate and move outward,
and calculate the corresponding Alfv\'en characteristic speeds from
Eq.~\eqref{eq:ffe_alfven_characteristic} using the local field
values at the $J_z$ extrema. The results show good agreement between
the measured and calculated speeds, indicating that the current branches
propagate outward as Alfv\'en modes.

\section{Fit to the dissipation rate}
\label{sec:dissipation_fit}

\begin{figure}[!t]
	\centering
	\includegraphics[width=0.92\columnwidth]{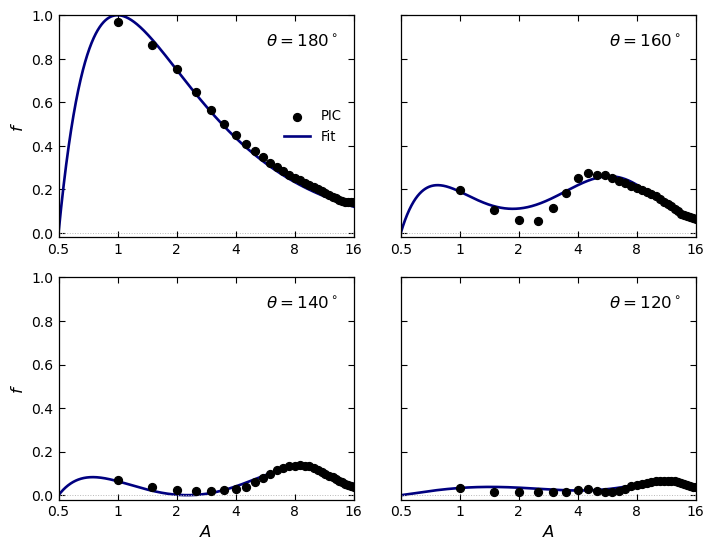}
	\caption{Comparison of the PIC dissipation rates (black points) with
	the empirical fit of Eq.~\eqref{eq:fit_formula} (blue curves) for
	representative polarization angles.}
	\label{fig:fit_diss}
\end{figure}

The dissipation rate is fitted by the following empirical form, which
provides a compact representation of the PIC results over the explored
parameter space, $0.5\leq A\leq16$ and
$90^\circ\leq\theta\leq180^\circ$:
\begin{equation}
\begin{aligned}
f(A,\theta)=\frac{1}{A^2}\Big[
&2A-1+xP_1(A)+x^4P_2(A) \\
&+xsP_3(A)+(1-s)P_4(A)
\Big],
\end{aligned}
\label{eq:fit_formula}
\end{equation}
where $x=\cos(\theta/2)$, $s=\sin(\theta/2)$, and
$P_k(A)=\sum_{n=0}^{6}c_{kn}A^n$.
The coefficients $c_{kn}$ are listed in
Table~\ref{tab:fit_coefficients}, and representative comparisons with
the PIC results are shown in Fig.~\ref{fig:fit_diss}.

\begin{table*}[t]
\caption{Coefficients $c_{kn}$ in the empirical fit
of Eq.~(\ref{eq:fit_formula}).}
\label{tab:fit_coefficients}
\centering
\begin{ruledtabular}
\begin{tabular}{crrrrrrr}
 & $c_{k0}$ & $c_{k1}$ & $c_{k2}$ & $c_{k3}$ &
 $c_{k4}$ & $c_{k5}$ & $c_{k6}$ \\
\hline
$P_1$ &
$16.57330$ & $272.70403$ & $-810.63048$ & $435.85727$ &
$-78.84804$ & $5.77447$ & $-0.14742$ \\
$P_2$ &
$-9.45286$ & $-95.89431$ & $310.15865$ & $-177.04055$ &
$33.06647$ & $-2.47144$ & $0.06396$ \\
$P_3$ &
$-13.83598$ & $-266.37151$ & $780.70705$ & $-422.16426$ &
$76.62382$ & $-5.62336$ & $0.14376$ \\
$P_4$ &
$-5.11542$ & $-127.92652$ & $358.87195$ & $-180.21239$ &
$31.29703$ & $-2.23028$ & $0.05588$
\end{tabular}
\end{ruledtabular}
\end{table*}

\end{document}